\documentclass{article}
\usepackage{spconf,amsmath,graphicx}
\usepackage{tikz}
\usepackage{comment}
\usepackage{amsmath}
\usepackage{hyperref}
\usepackage{fancyhdr}
\usetikzlibrary{shapes.geometric,arrows}

\title{On the Application of Log Compression and Enhanced Denoising in Contrast Enhancement of Digital Radiography Images}
\name{Asif M S and Mahesh Raveendranatha Panicker}
\address{Center for Computational Imaging and Department of Electrical Engineering, \\ Indian Institute of Technology Palakkad}
\begin{document}
%
\maketitle
\pagestyle{fancy}
\thispagestyle{fancy}
\fancyhead{}
\setlength{\headheight}{20pt}
\renewcommand{\headrulewidth}{0pt}
\fancyhead[L]{\fontsize{10}{10} \selectfont \textcolor{red}{This is an originally submitted version and has not been reviewed by independent peers. This work is licensed under a \href{https://creativecommons.org/licenses/by-nc-nd/4.0/}{\textcolor{red}{Creative Commons Attribution-NonCommercial-NoDerivatives (CC-BY-NC-ND) 4.0 License.}}}}
\fancyfoot{}
\begin{abstract}
Digital radiography (DR) is becoming popular for the point of care imaging in the recent past. To reduce the radiation exposure, controlled radiation based on as low as reasonably achievable (ALARA) principle is employed and this results in low contrast images.  To address this issue, post-processing algorithms such as the Multiscale Image Contrast Amplification (MUSICA) algorithm can be used to enhance the contrast of DR images even with a low radiation dose. In this study, a  modification of the MUSICA algorithm is investigated to determine the potential for further contrast improvement specifically for DR images. The conclusion is that combining log compression and its inverse at the appropriate stage with a multi-stage MUSICA and denoising is very promising. The proposed method resulted in an average of 66.5 \% increase in the mean contrast-to-noise ratio (CNR) for the test images considered.     
\end{abstract}


\section{Introduction}
\label{sec:intro}
One of the main issues with the digital radiography (DR) is the low contrast of the images which will eventually affect the diagnostics adaptability. Multi scale/frequency methods employing Laplacian pyramid and wavelets have been quite relevant for increasing the contrast of DR images \cite{vuylsteke1994multiscale} \cite{dippel2002multiscale}, where the former has been shown to result in less artefacts compared to the latter \cite{dippel2002multiscale}.

The multiscale image contrast amplification (MUSICA) algorithm, based on the laplacian pyramid concept, was introduced in  \cite{vuylsteke1994multiscale} and pioneered by AGFA healthcare \cite{agfa_musica}. In \cite{notohamiprodjo2022advances}, a qualitative comparison between the second generation MUSICA (MUSICA 2) and the third generation MUSICA (MUSICA 3) for various test images was presented, which depicts the importance of the algorithm in the radiography world. Improvements in the non-linear function used in conventional MUSICA have also been proposed in the recent past \cite{liu2022new}. 

In this work, a novel multi-stage MUSICA approach is introduced, capable of enhancing the contrast of the CT images much better than a single-stage MUSICA, with the aid of log compression and enhanced denoising. Though image enhancements by processing the image after taking log and then finally inverting the log has been done before \cite{zhang2006novel}, a multi-stage approach of MUSICA in the manner proposed here has not been explored before.   

\begin{figure*}[h]
    \centering
\includegraphics[width=\linewidth]{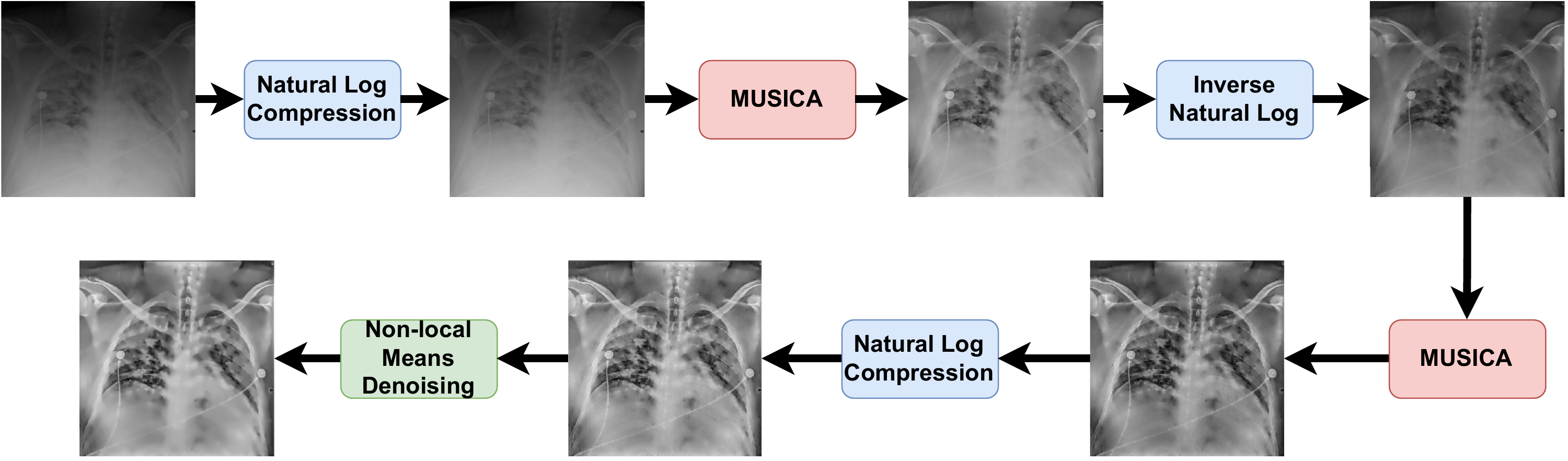}
\label{flow_diagram}
   \caption{Proposed framework which consists of applying MUSICA on a log compressed image followed by applying MUSICA on the enhanced inverselog image and non-local means denoising.}
    \label{fig:proposed_framework}
\end{figure*}

\section{Proposed Approach}
\label{sec:format}
\subsection{Review of MUSICA Algorithm}
The original MUSICA algorithm \cite{vuylsteke1994multiscale} involves decomposing the input image into a multi-resolution pyramid, known as the Laplacian pyramid, that represents local details at all levels, where each layer of the pyramid is termed as a detail layer. The detail layer coefficients are subsequently transformed by employing a nonlinear function as in \eqref{non_linear_eqn}. The non-linear function has the property of increasing the magnitude of low amplitudes and decreasing the magnitude of high amplitudes of the detail coefficients. The modified detail coefficients are reconstructed back resulting in a contrast-enhanced image. Unlike contrast enhancement methods that use sliding neighborhood techniques which often lead to the production of artifacts near sudden signal changes, such as at the boundary between bone and soft tissue, this algorithm is not based on sliding neighborhoods which reduces the chances of external artifacts appearing at high contrast edges and is widely applied to DR images of
chest, skull, spine, shoulder, pelvis, extremities and abdomen and finds application in non-destructive testing (NDT) industry also.  

\begin{equation}
\label{non_linear_eqn}
 \begin{array}{l} 
y(x)=\left\{\begin{array}{l}
\operatorname{aM} \frac{x}{x_c}\left(\frac{x_c}{M}\right)^p, \text { if }|x|<x_c \\
\operatorname{aM} \frac{x}{|x|}\left(\frac{|x|}{M}\right)^p, \text { if }|x|>x_c
\end{array}\right. \\
-M< x <M \text{ and } 0< x_c \ll M
\end{array}
\end{equation}

where $x$ are the detail coefficient values, $a$ is the global amplification factor (kept as $1$ in this work), $M$ are the upper bounds for the coefficient values. The key parameters which will affect the performance of the MUSICA algorithm are $x_c$ and $p$, which represents the noise control threshold and the non-linearity in the output. The noise control threshold $x_c$ if kept very low will amplify even granular details and hence increases the effective noise in the signal. The non-linearity factor $p$ is typically kept between 0.2 to 0.8 to achieve the desired result and higher the value less will be the non-linearity effect.
\subsection{Proposed Approach Employing Log Compression and Enhanced Denoising}
The proposed approach as shown in Fig. \ref{fig:proposed_framework} consists of the MUSICA algorithm applied at multiple stages. The input image, normalized to the range [0,1], is pre-processed with $f(X)$ as in \eqref{eq:fun_definition} (basically log compression) before doing contrast enhancement based on MUSICA. The choice of $ln(1+X)$ over $ln(X)$ is due to the fact 
that the logarithm expression is undefined for $X=0$.

\begin{equation}
\label{eq:fun_definition}
    f(X)= \ln (1+X) 
\end{equation}

 The purpose of $f(X)$ is to adjust the dynamic range of the input image such that the lower-intensity edges and the higher-intensity (dominating) edges are brought into the same range. After pre-processing the input image using log, the image is decomposed into aforementioned detail layers which are then modified using a nonlinear function \eqref{non_linear_eqn} as described in \cite{vuylsteke1994multiscale}. The image is then reconstructed as in the typical MUSICA algorithm \cite{vuylsteke1994multiscale}. The dynamic range is brought back to the original dynamic range by employing \eqref{eq:fun_inv_definition} after the MUSICA stage. 

\begin{equation}
\label{eq:fun_inv_definition}
    f^{-1}(X)= e^{X}-1 
\end{equation}

After the MUSICA on the log compressed image, the MUSICA is again repeated on the normal scale. At the output of this stage, noise (details) may be significantly enhanced. To reduce the noise, log compression is again employed followed by the non-local means denoising algorithm \cite{buades2005non}. This step is optional, where the degree of denoising can be controlled by changing the value of the hyperparameter $h$. 

 \subsection{Evaluation Metrics - Contrast to Noise Ratio (CNR)}
To evaluate the performance of the proposed approach, the contrast to noise ratio (CNR) is employed as in \cite{schaetzing2007agfa}. A CNR image is generated to quantify the contrast given for each coordinate $(i,j)$ of the image to be evaluated as in \eqref{eq:cnr_eqn}.       
 
 \begin{equation}
        \label{eq:cnr_eqn}
        CNR(i,j) = \frac{sdev_{3}(i,j)}{Noise_{3}}
\end{equation}

where, $sdev_{3}(i,j)$ and $Noise_{3}$ denotes the standard deviation image in the third detail layer corresponding to the pixel location $(i,j)$ and the reference noise level in that layer (computed by finding the pixel value corresponding to the maximum of the histogram of the standard deviation image, $sdev_{3}(i,j)$ ), respectively. The third detail layer is employed for the generation of the CNR image so that the contribution of the local noise can be reduced \cite{schaetzing2007agfa}. A $9\times9$ neighborhood window with a stride value of $1$ was used for the calculation of the standard deviation image. A white region in the CNR image corresponds to a higher CNR at that specific region. 

\begin{figure}
\centering
\includegraphics[page=1,width=\linewidth]{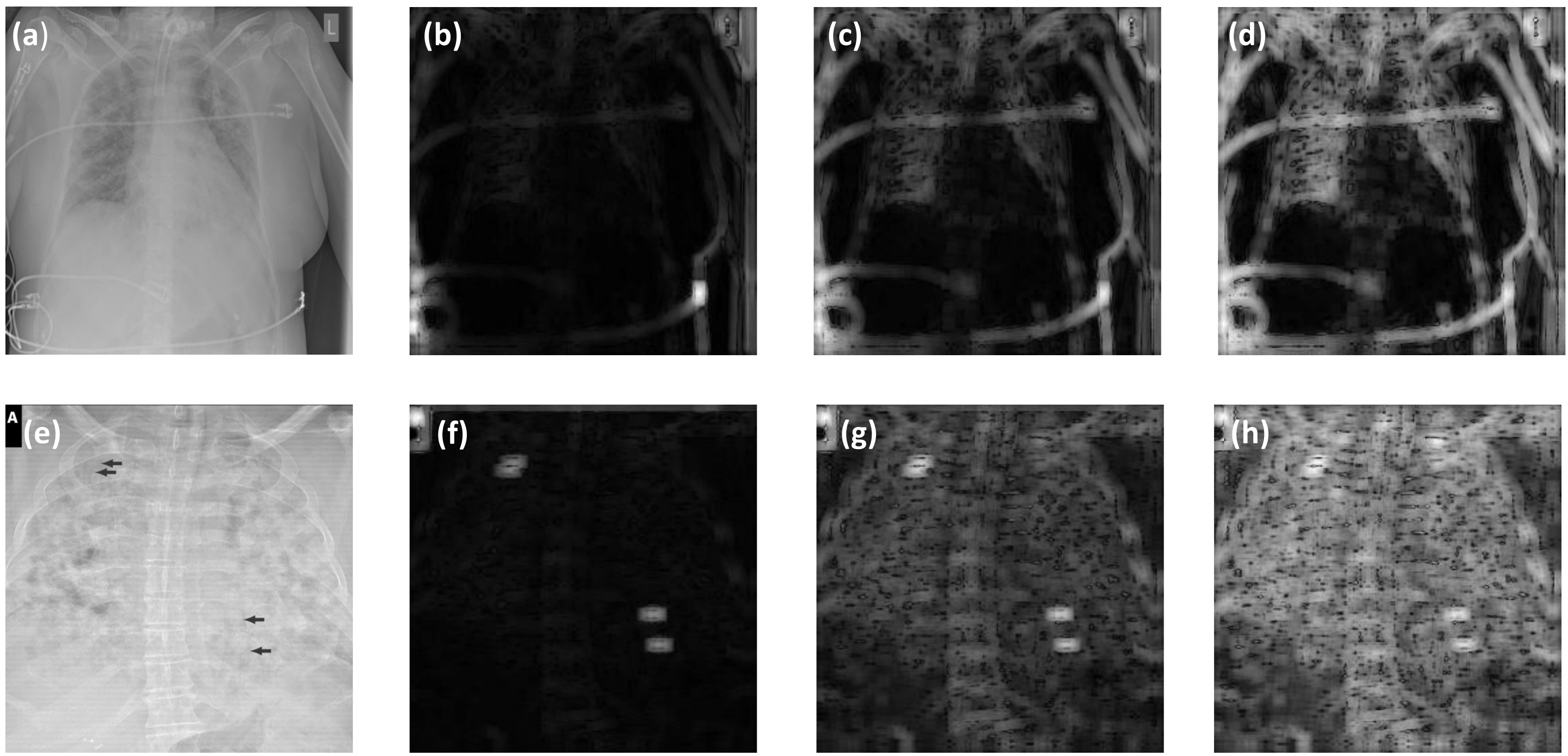}
\includegraphics[page=2,width=\linewidth]{CNR_Images.pdf}
\caption{Row-wise comparison among the original images (a,e,i,m) and the CNR images corresponding to the third detail layer for the original images (b,f,j,n), that of images processed using conventional MUSICA (c,g,k,o) and that of images processed using the proposed framework (d,h,l,p)}
\label{CNR_images}
\end{figure}

\section{Results}
\label{sec:results}
In this work, a set of $10$ DR images, corresponding to chest X-rays, were taken for the testing of the proposed algorithm from a publicly available dataset \cite{cohen2020covidProspective}. The CNR images using \eqref{eq:cnr_eqn} were generated for the original image, conventional MUSICA, and the proposed framework. A comparison of the aforementioned CNR images for four of the test images is given in Fig. \ref{CNR_images}. Each row corresponds to a unique test image. For the first test image shown (Fig. \ref{CNR_images} (a)), the CNR image (Fig. \ref{CNR_images} (b)) appears to be dark, which implies a low contrast. When the same test image is processed using conventional MUSICA, the resultant CNR image (Fig. \ref{CNR_images} (c)) is brighter which shows improvement in contrast. The CNR image of the test image processed using the proposed approach appears to improve the CNR (Fig. \ref{CNR_images} (d)), implying a significant contrast increase. A similar trend is observed for other test images (Fig. \ref{CNR_images} (e)-(h), (i)-(l), (m)-(p)) as well.  

\begin{figure}
 \centering
\includegraphics[page=1,width=\linewidth]{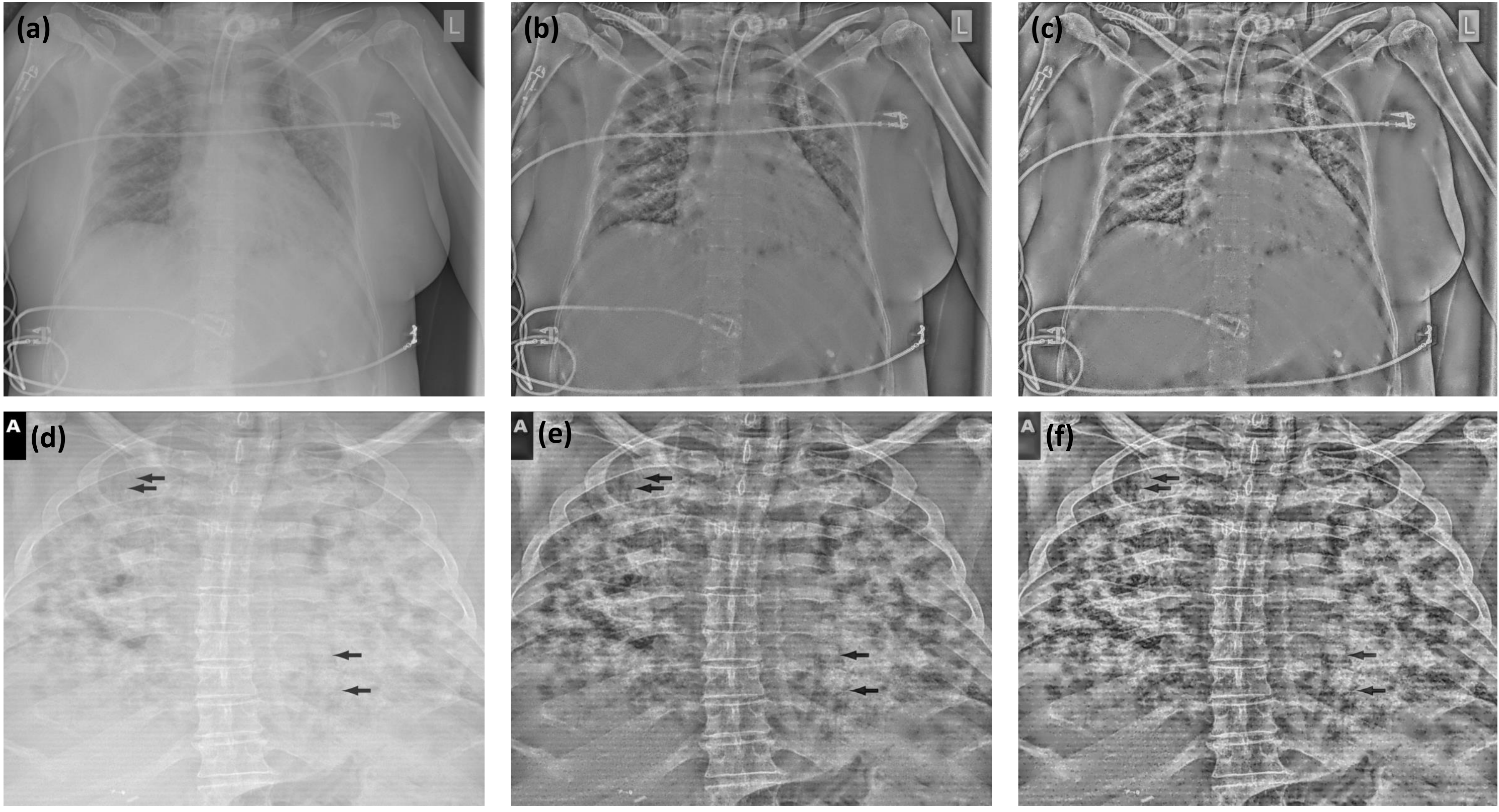}
\includegraphics[page=2,width=\linewidth]{Result_Images.pdf}
 \caption{Row-wise comparison among the original images (a,d,g,j), images processed using conventional MUSICA, (b,e,h,k) and that using the proposed framework (c,f,i,l)}
 \label{comparison_}
 \end{figure}

 The results of the proposed algorithm for four of the test images can be seen in Fig. \ref{comparison_}. For the first test image (Fig. \ref{comparison_} (a)), the conventional MUSICA algorithm results in a contrast-enhanced image as can be qualitatively observed in Fig. \ref{comparison_} (b). For the same image, the proposed approach is observed to result in a better contrast enhanced image (Fig. \ref{comparison_} (c)). A similar observation holds true for the other test images as in Fig. \ref{comparison_} (d)-(f), (g)-(i) and (j)-(l). The number of levels of decomposition was chosen as $7$ as it is stated to provide a clear separation between anatomical structures and diagnostic details \cite{hoeppner2002equalized}. The non-linearity parameter $p$ was taken as $0.5$ at all the decomposition levels. The values for $M$ and $a$ chosen were both $1$ and $x_c$ was chosen as $0.01$.  A patch size of $7$ and a search window size of $21$ has been used while employing the non-local means denoising algorithm.

\begin{figure*}[t]
 \centering
\includegraphics[width=0.7\linewidth]{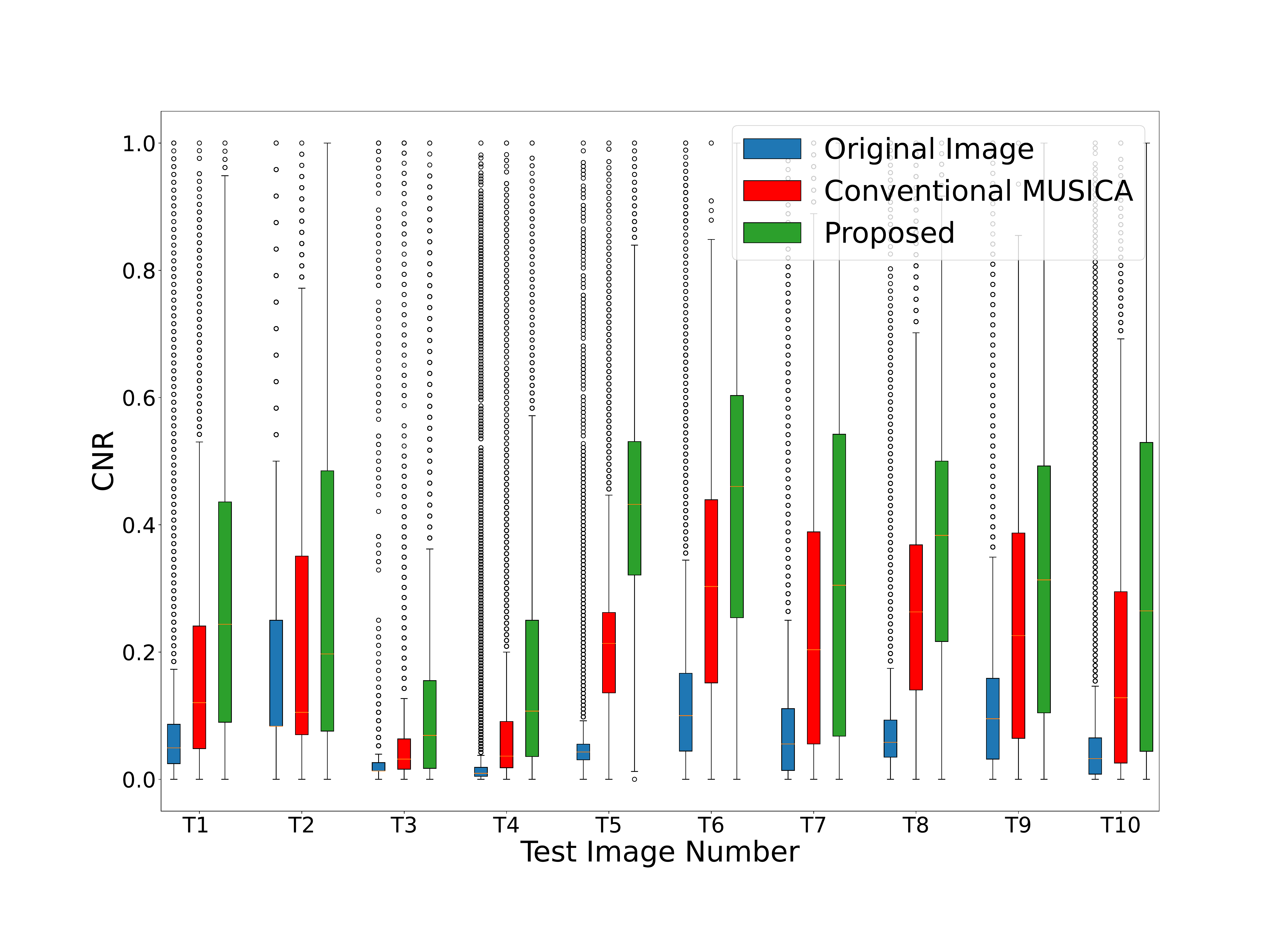}
 \caption{A box plot for the CNR Image comparison among Original Image, Conventional MUSICA and the Proposed algorithm}
 \label{cnr_box_plot}
 \end{figure*}

The proposed algorithm showed much better promise in enhancing the contrast of the original image than that of conventional MUSICA. The improvement in the CNR is further quantified in Fig. \ref{cnr_box_plot}. The CNR images similar to as depicted in Fig. \ref{CNR_images} was taken for all the $10$ test images to obtain the box plot as shown in  Fig. \ref{cnr_box_plot}. The plot shows a consistent increase in the median CNR value across all the ten test images when processed using the proposed approach rather than when the conventional MUSICA algorithm is employed. It is also interesting to note that the standard deviation of the CNR is also much better for the proposed framework and this shows that the contrast ratio (highest to lowest contrast) is also higher for the proposed approach. 
The algorithm was implemented on an Intel(R) Core (TM) i5-8265U CPU (1.60 GHz) PC and takes an average of $23s$ of compute time for the $10$ test images. The implementation of the proposed framework has been made open-source and is available at \cite{Image_Enhancement}. In the proposed approach, the parameters such as $p$ and $x_c$ have been assumed constant. The future work involves an accelerated optimization framework which will give the optimal values of the above parameters such that the CNR is maximized.
\section{Conclusions}
A multi-stage MUSICA approach for enhancing the contrast of DR images has been proposed in this work. The proposed approach employs the MUSICA on both the log compressed image and the image at normal scale, along  with non-local means based denoising.  The results  show that the proposed algorithm has enhanced the contrast of the input DR images, resulting in higher CNR values ($\sim$ 0.30 $\pm$ 0.20) compared to conventional MUSICA ($\sim$ 0.19 $\pm$ 0.15). 


\bibliographystyle{IEEEbib}
\bibliography{Template}

\end{document}